\documentclass[aps,preprintnumbers,prb,showpacs,superscriptaddress]{revtex4}
\usepackage{mathrsfs}
\usepackage{amsfonts}
\usepackage{graphicx}
\usepackage{amsmath}
\usepackage{amssymb}

\begin{document}
\title{Some exact identities connecting one- and two-particle Green's functions in spin-orbit coupling systems}
\author{Hua-Tong Yang}
\email{htyang@pku.edu.cn}\affiliation{School of Physics,Peking
University, Beijing 100871,China}

\begin{abstract}
Some exact identities connecting the one- and two-particle Green's
functions in the presence of spin-orbit coupling have been derived.
These identities is similar to the usual Ward identity in the
particle or charge transport theory and a satisfying spin transport
theory in spin-orbit coupling system should also preserve these
identities.
\end{abstract}
\pacs{72.25.Dc, 72.25.Ba, 72.25.Mk}
\maketitle

\section{Introduction}
In usual quantum transport theory of charge (or particles), the
derivatives of exact single- and two-particle Green's functions(or
equivalent vertexes) satisfy an exact identity, called generalized
Ward identity,\cite{Strinati} which essentially represents the
conservation law or gauge invariance. On the other hand, in a
practical calculation of the quantum transport phenomena we often
adopt some approximation methods. Therefore, it is a very important
criterion to ensure that the approximation remain this exact
identity, this kind of approximations is called as self-consistent
or conserving approximation.\cite{Baym&Kadanoff} It turns out that
the self-energy must be taken as a $\Phi$-derivative
form.\cite{Baym-pr1962} Recently, there has been increasing interest
in spin transport in spin-orbit coupling systems and some resulting
effects, e.g., spin-Hall effect(SHE),\cite{Zhang,
Sinova,Inoue,Dimitrova,Mishchenko,Schliemann&Loss-69,Nomura,Chalaev
and Loss, Raimondi and Schwab, Khaetskii, Schliemann&Loss-71, Liu &
Lei, Bryksin} because it maybe provide a promising method to
manipulate the electronic spin without application of magnetic
field\cite{Kato-Nature, Kato, Wunderlich, Koga} in
spintronics.\cite{Rashba-0309441,Zutic et al. rmp 76} As well known,
the spin is not a conserved quantity in the presence of spin-orbit
coupling, therefore, a natural question is that whether there are
also some exact identities similar to the usual Ward identity in the
charge transport theory? In this Letter we will derive some exact
identities connecting the one- and two-body Green's functions in the
presence of spin-orbit coupling. It is a consequence of a more
general continuity-like equation, which was given as an equation
valid for average values of the spin-density, current and torque in
our earlier paper,\cite{Yang-Liu} here we shall prove that it is
valid not only for the averages of these quantities, but also holds
as an operator identity. Moreover, it will imply some other exact
identities connecting various correlation functions, which represent
more general matrix elements of the spin-density, current and
torque. Therefore, a self-consistent or conserving approximation for
the response of the spin- density, current and torque must also
preserve these identities.

\section{The main identities}
\subsection{Model and Hamiltonian} Without loss of generality, the
Hamiltonian with a general spin-orbit coupling can be written as
\begin{eqnarray}\label{Hamiltonian}
\hat{\bf H}&=&\int \psi_\alpha^{\dag}({\bf x})[\hat{h}_0({\bf x},
\nabla)]_{\alpha\beta}\psi_{\beta}({\bf x})+ \int v({\bf x}-{\bf
x}')\psi_{\alpha}^{\dag}({\bf x})\psi_{\beta}^{\dag}({\bf
x}')\psi_{\beta}({\bf x}')\psi_{\alpha}({\bf x}).
\end{eqnarray} Here and hereafter a summation over repeated suffix is
implied and
\begin{eqnarray}\label{one-body Hamiltonian}
\hat{h}_0=\frac{1}{2m}(-i\widetilde{\nabla})^2
+e\phi+\hat{h}_{\textmd{so}}\end{eqnarray} with
$\widetilde{\nabla}_i\equiv
\widetilde{\partial}_i={\partial}_i-i\frac{e}{c}A_{i}$, $i=x,y,z,$
$\hat{h}_{\textmd{so}}$ is the spin-orbit coupling term. We suppose
\begin{eqnarray}
\hat{h}_{\textmd{so}}=\hat{H}^{(0)}+\hat{H}^{(1)}_{i}(-i\widetilde{\partial}_i)+\hat{H}^{(2)}_{ij}
(-i\widetilde{\partial}_i)(-i\widetilde{\partial}_j)+\hat{H}^{(3)}_{ijk}(-i\widetilde{\partial}_i)
(-i\widetilde{\partial}_j)(-i\widetilde{\partial}_k)+\cdots,\label{so-coupling}
\end{eqnarray}  with $\hat{H}^{(n)}_{i_1\cdots i_n}$ $(n=0,1,2,\cdots)$
an operator acting on spin variables, meanwhile, it is also the
component of a tensor of rank $n$'th and is supposed to be symmetric
under the permutations of its suffixes $i_1,i_2,\cdots, i_n$.
Moreover, we assume that all $\hat{H}^{(n)}_{i_1\cdots i_n}$ are
independent of ${\bf x}$ except $\hat{H}^{(0)}$.

\subsection{Two Mathematical Lemmas}
Firstly, we have to introduce two useful identities. The first one
can be immediately verified by using the technique due to Takahashi,
Umezawa and Lurie.\cite{Takahashi}

\textbf{Lemma.1} Let $$\Psi({\bf x})=\left( \begin{array}{c}\psi_1({\bf x})\\ \psi_2({\bf x})\\
\vdots\\ \psi_N({\bf x})\end{array}\right),~~\Phi^{\dag}({\bf
x})=\left(\phi_1^{\dag}({\bf x}), \phi_2^{\dag}({\bf x}), \cdots,
\phi_N^{\dag}({\bf x}) \right)$$ are functions of ${\bf x}$. ${\bf
A}_{i_1i_2\cdots i_n}$ is a matrix of $N\times N$, and is assumed to
be symmetric with respect to the permutations of its suffixes. It is
easy to verify that for any integer $n\geq1$, we have
\begin{eqnarray}\label{Lemma1}
\Phi^{\dag}{\bf A}_{i_1i_2\cdots i_n}(\partial^n_{i_1 i_2\cdots i_n}
\Psi)-(-1)^{n} (\partial^n_{i_1 i_2\cdots i_n}\Phi^{\dag}){\bf
A}_{i_1i_2\cdots i_n}\Psi =\partial_i J_i({\bf x}) =\nabla\cdot{\bf
J}({\bf x}),
\end{eqnarray} where
\begin{eqnarray}
J_i({\bf x}) &=&\Phi^{\dag} {\bf A}_{i,i_1\cdots
i_{n-1}}(\partial^{n-1}_{i_1\cdots i_{n-1}}
\Psi)+(-\partial_{i_1}\Phi^{\dag}) {\bf A}_{i,i_1\cdots
i_{n-1}}(\partial^{n-2}_{i_2\cdots i_{n-1}}\Psi)+\cdots+(-1)^{n-1}
(\partial^{n-1}_{i_1\cdots i_{n-1}} \Phi^{\dag}){\bf A}_{i,i_1\cdots
i_{n-1}}\Psi\nonumber\\
&=&\sum_{k=0}^{n-1}(-1)^k(\partial^{k}_{i_1\cdots i_{k}}\Phi^{\dag})
{\bf A}_{i,i_1\cdots i_{n-1}}(\partial^{n-k-1}_{i_{k+1}\cdots
i_{n-1}}\Psi)
=\sum_{k=0}^{n-1}(-1)^k\Phi^{\dag}\overleftarrow{\partial}^{k}_{i_1\cdots
i_{k}}{\bf A}_{i,i_1\cdots
i_{n-1}}\overrightarrow{\partial}^{n-k-1}_{i_{k+1}\cdots
i_{n-1}}\Psi.
\end{eqnarray}
Here and hereafter we use $\overrightarrow{\partial}$ to indicate
the differentiation acting on its right-hand side factor $\Psi$,
while $\overleftarrow{\partial}$ acts on its left-hand side factor
$\Psi^{\dag}$.

Moreover, we can verify the following theorem:

\textbf{Lemma.2} Let $n\geq1$ and
$\overline{\partial}_i\equiv\frac{1}{2}(\overrightarrow{\partial}_i-\overleftarrow{\partial}_i)$,
we have
\begin{eqnarray}\label{Lemma2}
\Phi^{\dag}{\bf A}_{i_1i_2\cdots i_n}(\partial^n_{i_1 i_2\cdots i_n}
\Psi)-\Phi^{\dag}{\bf A}_{i_1i_2\cdots
i_n}\overline{\partial}^{n}_{i_1 i_2\cdots i_n}\Psi =\partial_i
J'_i({\bf x})=\nabla\cdot{\bf J}'({\bf x}),\nonumber\\
(-1)^n(\partial^n_{i_1 i_2\cdots i_n}\Phi^{\dag}){\bf
A}_{i_1i_2\cdots i_n}\Psi-\Phi^{\dag}{\bf A}_{i_1i_2\cdots
i_n}\overline{\partial}^{n}_{i_1 i_2\cdots i_n}\Psi =\partial_i
J''_i({\bf x})=\nabla\cdot{\bf J}''({\bf x}),
\end{eqnarray} where
\begin{eqnarray}\label{J1-P}
J'_i({\bf x})&=&\frac{1}{2}\left(\sum_{k=0}^{n-1}\Phi^{\dag}{\bf
A}_{i,i_1\cdots i_{n-1}}\overline{\partial}^{k}_{i_1\cdots i_k
}\overrightarrow{\partial}^{n-k-1}_{i_{k+1}\cdots
i_{n-1}}\Psi\right)\equiv\frac{1}{2}P'_i({\bf x}),\nonumber\\
J''_i({\bf
x})&=&-\frac{1}{2}\left(\sum_{k=0}^{n-1}(-1)^{k}\Phi^{\dag}{\bf
A}_{i,i_1\cdots i_{n-1}}\overleftarrow{\partial}^{k}_{i_1\cdots i_k
}\overline{\partial}^{n-k-1}_{i_{k+1}\cdots
i_{n-1}}\Psi\right)\equiv\frac{1}{2}P''_i({\bf x}),
\end{eqnarray}
and $\overline{\partial}^{n}_{i_1 i_2\cdots i_n}=
\frac{1}{2}(\overrightarrow{\partial}_{i_1}-\overleftarrow{\partial}_{i_1})\cdots
\frac{1}{2}(\overrightarrow{\partial}_{i_n}-\overleftarrow{\partial}_{i_n})$.

\textbf{Proof.} From the Def.(\ref{J1-P}), we can directly verify
that
\begin{eqnarray}
\partial_i J'_i({\bf x})=\frac{1}{2}\left(\overrightarrow{\partial}_i
+\overleftarrow{\partial}_i\right)P'_i({\bf x})
=(\overrightarrow{\partial}_i-\overline{\partial}_i)P_i({\bf x})
=\Phi^{\dag}{\bf A}_{i_1i_2\cdots
i_n}\left(\overrightarrow{\partial}^n_{i_1 i_2\cdots i_n}
-\overline{\partial}^{n}_{i_1 i_2\cdots i_n}\right)\Psi
\end{eqnarray}
The second line of Eq.(\ref{Lemma2}) can be verified in a same way.
$\Box$

The Eq.(\ref{Lemma1}) and (\ref{Lemma2}) also remain valid if all
$\partial_i$'s are replaced by $\widetilde{\partial}_i$'s because
$\overleftarrow{\widetilde{\partial}_i}=\overleftarrow{\partial}_i+i\frac{e}{c}A_{i}$,
so $\overrightarrow{\partial}_i+\overleftarrow{\partial}_i=
\overrightarrow{\widetilde{\partial}}_i+\overleftarrow{\widetilde{\partial}}_i$.

\subsection{Generalized equation of continuity}
Now we consider a density operator $\rho_{\sigma^\mu}({\bf
x})=\psi_\alpha^{\dag}({\bf
x})\sigma^\mu_{\alpha\beta}\psi_\beta({\bf x})=\Psi^{\dag}({\bf
x})\sigma^\mu\Psi({\bf x})$, where
$$\Psi^{\dag}=(\psi^{\dag}_{\uparrow},\psi^{\dag}_{\downarrow}),~~
\Psi=\left(\begin{array}{c}
\psi_{\uparrow}\\
\psi_{\downarrow}\end{array}\right ).$$  $\rho_{\sigma^\mu}$ will
represent the particle density if the matrix $\sigma^\mu$ is a unit
matrix $\sigma^0=\left(\begin{array}{cc}1&0\\0&1\end{array}
\right)$, or the spin density if $\sigma^\mu$ equals to the Pauli
matrix $\sigma^1$, $\sigma^2$, $\sigma^3$. According to the equation
of motion
\begin{eqnarray}
\frac{\partial }{\partial t}\Psi^{\dag}({\bf
x}t)&=&i\left[\Psi^{\dag}({\bf x}t)\hat{h}_0({\bf
x},-\overleftarrow{\partial})+\int v({\bf x}-\bar{\bf
x})\Psi^{\dag}({\bf x}t) \Psi^{\dag}(\bar{\bf x}t)\Psi(\bar{\bf x}t)
\right],\nonumber\\
\frac{\partial}{\partial t}\Psi({\bf x}t)&=&-i\left[\hat{h}_0({\bf
x},\overrightarrow{\partial})\Psi({\bf x}t)+\int v({\bf x}-\bar{\bf
x})\Psi^{\dag}(\bar{\bf x}t)\Psi(\bar{\bf x}t) \Psi({\bf
x}t)\right],
\end{eqnarray} where the bar over $\bar{\bf x}$ is to indicate the integral
variable, we have
\begin{eqnarray}\label{eq 10}\frac{\partial}{\partial t}\rho_{\sigma^\mu}({\bf
x})= -i\Psi^{\dag}({\bf x}t)\left[\sigma^\mu\hat{h}_0({\bf
x},\overrightarrow{\partial})-\hat{h}_0({\bf
x},-\overleftarrow{\partial})\sigma^\mu\right]\Psi({\bf x}t).
\end{eqnarray} The four-operators terms are canceled because $$\int[v({\bf x}-\bar{\bf
x})-v({\bf x}-\bar{\bf x})] \Psi^{\dag}({\bf
x}t)\Psi^{\dag}(\bar{\bf x}t)\Psi(\bar{\bf x}t) \Psi({\bf x}t)=0,$$
Substitute Eq.(\ref{one-body Hamiltonian}) into Eq.(\ref{eq 10}) we
obtain
\begin{eqnarray}\label{eqmotion}\frac{\partial}{\partial t}\rho_{\sigma^\mu}({\bf
x})=-i\Psi^{\dag}({\bf
x}t)\left\{\frac{1}{2m}\left(\overleftarrow{\Delta}^2
\sigma^\mu-\sigma^\mu\overrightarrow{\nabla}^2\right)+\left[\sigma^\mu\hat{h}_{\textmd{so}}({\bf
x},\overrightarrow{\partial})-\hat{h}_{\textmd{so}}({\bf
x},-\overleftarrow{\partial})\sigma^\mu\right]\right\}\Psi({\bf
x}t).
\end{eqnarray}
The right-hand side of the Eq.(\ref{eqmotion}) includes two
different kinds of terms. The first one comes from the kinetic
energy
\begin{eqnarray}\frac{-i}{2m}\Psi^{\dag}({\bf x}t)\left[\overleftarrow{\Delta}^2
\sigma^\mu-\sigma^\mu\overrightarrow{\nabla}^2\right]\Psi({\bf
x}t)=-\nabla \cdot{\bf J}^{\sigma^\mu}_0,
\end{eqnarray} where
\begin{eqnarray}\label{current-0}
{\bf J}^{\sigma^{\mu}}_0({\bf
x}t)=\frac{-i}{2m}\left[\Psi^{\dag}({\bf x}t)
\sigma^\mu\nabla\Psi({\bf x}t)-\nabla\Psi^{\dag}({\bf x}t)
\sigma^\mu\Psi({\bf x}t)\right ].
\end{eqnarray}
The second term is due to the spin-orbit coupling
\begin{eqnarray}\label{s-o term}
-i\Psi^{\dag}\left[\sigma^\mu\hat{h}_{\textmd{so}}({\bf
x},\overrightarrow{\partial})-\hat{h}_{\textmd{so}}({\bf
x},-\overleftarrow{\partial})\sigma^\mu\right]\Psi =-\sum_n
(-i)^{n-1}\Psi^{\dag}[\sigma^\mu\hat{H}^{(n)}_{i_1\cdots
i_n}\overrightarrow{\partial}^{n}_{i_1\cdots
i_n}-(-1)^n\hat{H}^{(n)}_{i_1\cdots
i_n}\sigma^\mu\overleftarrow{\partial}^{n}_{i_1\cdots i_n}]\Psi.
\end{eqnarray}
Using the identities
\begin{eqnarray}
\sigma^\mu\hat{H}^{(n)}_{i_1\cdots
i_n}&=&\frac{1}{2}\{\sigma^\mu,\hat{H}^{(n)}_{i_1\cdots
i_n}\}+\frac{1}{2}[\sigma^\mu, \hat{H}^{(n)}_{i_1\cdots
i_n}],\nonumber\\
\hat{H}^{(n)}_{i_1\cdots
i_n}\sigma^\mu&=&\frac{1}{2}\{\sigma^\mu,\hat{H}^{(n)}_{i_1\cdots
i_n}\}-\frac{1}{2}[\sigma^\mu, \hat{H}^{(n)}_{i_1\cdots
i_n}],\end{eqnarray} Eq.(\ref{s-o term}) becomes
\begin{eqnarray}\label{s-o coupling term}
i\Psi^{\dag}\left[\sigma^\mu\hat{h}_{\textmd{so}}({\bf
x},\overrightarrow{\partial})-\hat{h}_{\textmd{so}}({\bf
x},-\overleftarrow{\partial})\sigma^\mu\right]\Psi
&=&\sum_n\frac{(-i)^{n-1}}{2}\Psi^{\dag}[\{\sigma^{\mu},\hat{H}^{(n)}_{i_1\cdots
i_n}\}(\overrightarrow{\partial}^{n}_{i_1\cdots
i_n}-(-1)^n\overleftarrow{\partial}^{n}_{i_1\cdots
i_n})\nonumber\\&+& [\sigma^{\mu}, \hat{H}^{(n)}_{i_1\cdots
i_n}](\overrightarrow{\partial}^{n}_{i_1\cdots
i_n}+(-1)^n\overleftarrow{\partial}^{n}_{i_1\cdots i_n})]\Psi.
\end{eqnarray}
Let $\Phi^{\dag}=\Psi^{\dag}$, ${\bf A}_{i_1\cdots
i_n}=\{\hat{H}^{(n)}_{i_1\cdots i_n},\sigma^{\mu}\}$ and use the
Lemma.1, the first term of Eq.(\ref{s-o coupling term}) can be
written as $\nabla\cdot{\bf J}^{\sigma^{\mu}}_\textmd{so}$, where
\begin{eqnarray}\label{J-SO}
(J^{\sigma^{\mu}}_{\textmd{so}})_i=\sum_n
\frac{(-i)^{n-1}}{2}\Psi^{\dag}\{\sigma^{\mu},\hat{H}^{(n)}_{i_1i_2\cdots
i_n}\}\sum_{k=0}^{n-1}(-1)^k\overleftarrow{\partial}^{k}_{i_2\cdots
i_{k-1}}\overrightarrow{\partial}^{n-k-1}_{i_{k}\cdots i_n}\Psi.
\end{eqnarray}
The second term of Eq.(\ref{s-o coupling term}) can be rewritten as
$\nabla\cdot\tilde{{\bf
J}}^{\sigma^{\mu}}_\textmd{so}-T^{\sigma^{\mu}}$, where the
$T^{\sigma^{\mu}}$ and $\tilde{{\bf J}}^{\sigma^{\mu}}$ are
respectively defined as
\begin{eqnarray}\label{Torque}
T^{\sigma^{\mu}}=-\sum_n(-i)^{n-1}\Psi^{\dag} \left[\sigma^{\mu},
\hat{H}^{(n)}_{i_1\cdots i_n}\right]\overline{
\partial}^{n}_{i_1\cdots i_n}\Psi.
\end{eqnarray}
\begin{eqnarray}\label{new-current}
(\tilde{J}_{\textmd{so}}^{\sigma^{\mu}})_i=\sum_{n\geq2}\frac{(-i)^{n-1}}{4}\Psi^{\dag}
\left[\sigma^{\mu}, \hat{H}^{(n)}_{i,i_1\cdots
i_{n-1}}\right]\sum_{k=0}^{n-1}(\overrightarrow{\partial}^{k}_{i_1\cdots
i_k}-(-1)^k \overleftarrow{\partial}^{k}_{i_1\cdots i_k})\overline{
\partial}^{n-k-1}_{i_{k+1}\cdots i_{n-1}}\Psi.
\end{eqnarray}
We notice that the $T^{\sigma^{\mu}}$ is given by a polynomial of
$\overline{\partial}_i=1/2(\overrightarrow{\partial}_i-\overleftarrow{\partial}_i)$,
therefore, it can not be rewritten as a divergence
form.\cite{Yang-Liu} Therefore, we obtain the following identity
\begin{eqnarray}\label{cont-equation}\frac{\partial \rho_{\sigma^\mu}({\bf x}t)}{\partial
t}+\nabla\cdot \left[{\bf J}^{\sigma^{\mu}}_0({\bf x}t)+{\bf
J}^{\sigma^{\mu}}_{\textmd{so}}({\bf x}t)+\tilde{\bf
J}^{\sigma^{\mu}}_{\textmd{so}}({\bf
x}t)\right]=T^{\sigma^{\mu}}({\bf x}t).\end{eqnarray}  The
generalized continuity equation given in our previous paper
\cite{Yang-Liu} can be obtained by takeing the average value on both
sides of Eq.(\ref{cont-equation}). Evidently,
Eq.(\ref{cont-equation}) includes two different cases: 1).
${\sigma^\mu}={\sigma^0}$, in this case $\rho_{\sigma^{0}}$ is the
number density $\rho$, from the Def.(\ref{Torque}) and
Def.(\ref{new-current}) we have $T^{\sigma^{0}}=0$, $\tilde{\bf
J}_{\textmd{so}}^{\sigma^{0}}=0$, so, ${\bf J}^{\sigma^{0}}+{\bf
J}^{\sigma^{0}}_{\textmd{so}}$ is the current density ${\bf J}$  and
we have the following local conservation law of number
\begin{eqnarray}\frac{\partial \rho({\bf x}t)}{\partial
t}+\nabla\cdot \left({\bf J}_0({\bf x}t)+{\bf J}_{\textmd{so}}({\bf
x}t)\right)=0.\end{eqnarray} 2).
${\sigma^\mu}={\sigma^1},{\sigma^2}, {\sigma^3}$,
Eq.(\ref{cont-equation}) represents the continuity-like equation of
spin.
\subsection{Exact identities connecting the correlation functions}
Because Eq.(\ref{cont-equation}) is an operator identity, it is
valid for arbitrary matrix-element of both sides of this equation.
Therefore, for any operator $O({\bf x}'t')$, we have the following
identity:
\begin{eqnarray}
&&\frac{\partial}{\partial t}\langle\mathcal
{T}[\rho_{\sigma^\mu}({\bf x}t)O({\bf x}'t')]\rangle+\nabla\cdot
\langle\mathcal {T}[({\bf J}^{\sigma^{\mu}}_0({\bf x}t)+{\bf
J}^{\sigma^{\mu}}_{\textmd{so}}({\bf x}t)+\tilde{\bf
J}^{\sigma^{\mu}}_{\textmd{so}}({\bf x}t))O({\bf
x}'t')]\rangle\nonumber\\
&&=\langle\mathcal {T}[T^{\sigma^{\mu}}({\bf x}t)O({\bf
x}'t')]\rangle+\delta(t-t')\langle[\rho_{\sigma^\mu}({\bf
x}t),O({\bf x}'t')]\rangle,\end{eqnarray} where the $\mathcal {T}$
is the time-ordering operator and $O$ may equal to, e.g.,
$\rho_{\sigma^\mu}$ or ${\bf J}^{\sigma^\mu}$ etc. Similarly, for
arbitrary operators $A$ and $B$, we have
\begin{eqnarray}\label{2-operator-correlation}
\frac{\partial}{\partial t_1}\langle\mathcal
{T}\{\rho_{\sigma^\mu}(1)A(2)B(3)\}\rangle+\nabla_1\cdot
\langle\mathcal {T}\{[{\bf J}^{\sigma^{\mu}}_0(1)+{\bf
J}^{\sigma^{\mu}}_{\textmd{so}}(1)+\tilde{\bf
J}^{\sigma^{\mu}}_{\textmd{so}}(1)]A(2)B(3)\}\rangle-\langle\mathcal
{T}\{T^{\sigma^{\mu}}(1)A(2)B(3)\}\rangle\nonumber\\
=\delta(t_1-t_2)\langle\mathcal
{T}\{[\rho_{\sigma^\mu}(x_1t_1),A(x_2t_1)]B(3)\}\rangle+\delta(t_1-t_3)\langle\mathcal
{T}\{A(2)[\rho_{\sigma^\mu}(x_1 t_1),B(x_3
t_1)]\}\rangle.\end{eqnarray} For example, if
$A(2)=\psi_{\alpha}(2)$ and $B(3)=\psi_{\beta}^{\dag}(3)$,
Eq.(\ref{2-operator-correlation}) becomes
\begin{eqnarray}\label{identity}
&&\frac{\partial}{\partial t_1}\langle\mathcal
{T}\{\rho_{\sigma^\mu}(1)\psi_{\alpha}(2)\psi_{\beta}^{\dag}(3)\}\rangle+\nabla_1\cdot
\langle\mathcal {T}\{[{\bf J}^{\sigma^{\mu}}_0(1)+{\bf
J}^{\sigma^{\mu}}_{\textmd{so}}(1)+\tilde{\bf
J}^{\sigma^{\mu}}_{\textmd{so}}(1)]\psi_{\alpha}(2)\psi_{\beta}^{\dag}(3)\}\rangle
\nonumber\\&&=\langle\mathcal
{T}\{T^{\sigma^{\mu}}(1)\psi_{\alpha}(2)\psi_{\beta}^{\dag}(3)\}\rangle
-\delta(1-2)\sigma^{\mu}_{\alpha\nu}\langle\mathcal
{T}\{\psi_{\nu}(2)\psi_{\beta}^{\dag}(3)\}\rangle+\delta(1-3)\langle\mathcal
{T}\{\psi_{\alpha}(2)\psi_{\nu}^{\dag}(3)]\}\rangle\sigma_{\nu\beta}^{\mu}\nonumber\\
&&=\langle\mathcal
{T}\{T^{\sigma^{\mu}}(1)\psi_{\alpha}(2)\psi_{\beta}^{\dag}(3)\}\rangle
-i\left[\delta(1-2)\sigma^{\mu}G(2,3)-\delta(1-3)G(2,3)\sigma^{\mu}\right]_{\alpha\beta}.\end{eqnarray}

These identities represents the exact relations satisfied by the
generalized responses functions of the spin- density, current and
spin torque to some external perturbations. The physical essential
of these identities is that the responses of the spin-density,
current and torque in non-equilibrium should also obey the
generalized continuity equation, since this equation is an exact
operator identity, which would be satisfied in any situation.
Therefore, if we calculate the spin-density, current and torque by
some approximate methods, then the approximation must also preserve
these exact identities.
\subsection{Rashba Model }
Because the expressions of the spin-current and torque are dependent
on the concrete Hamiltonian, therefore, the form of these identities
also depend on Hamiltonian of the system. As an example, we consider
a 2D system with Rashba spin-orbit coupling
\begin{eqnarray}h_R=\alpha(\sigma_xp_y-\sigma_yp_x).\end{eqnarray}
According to the definition of Def.(\ref{current-0}), (\ref{J-SO})
and (\ref{Torque}), we have
\begin{eqnarray}\label{correlation}&&\langle\mathcal
{T}[\rho_{\sigma^{\mu}}(1)\psi_{\nu}(2)\psi_{\nu'}^{\dag}(3)]\rangle
=\sigma^{\mu}_{\eta'\eta}G_{\eta\nu\eta'\nu'}(1,2;1^+,3),\nonumber\\
&&\langle \mathcal {T}[{\bf
J}_0^{\sigma^{\mu}}(1)\psi_{\nu}(2)\psi_{\nu'}^{\dag}(3)]\rangle
=\frac{1}{2im}\sigma^{\mu}_{\eta'\eta}\left[\left
(\nabla_1-\nabla_{1'}\right)G_{\eta\nu\eta'\nu'}(1,2;1',3)\right]_{1'=1^+},\nonumber\\
&&\langle\mathcal {T}[{\bf
J}_{\textmd{so}}^{\sigma^{\mu}}(1)\psi_{\nu}(2)\psi_{\nu'}^{\dag}(3)]\rangle
=\frac{\alpha}{2}\left(-\{\sigma^\mu, \sigma^y \}_{\eta'\eta},
\{\sigma^\mu, \sigma^x \}_{\eta'\eta}
\right)G_{\eta\nu\eta'\nu'}(1,2;1^+,3),\nonumber\\
&&\langle \mathcal
{T}[T^{\sigma^{\mu}}(1)\psi_{\nu}(2)\psi_{\nu'}^{\dag}(3)]\rangle
=\frac{\alpha}{2}\left\{([\sigma^\mu, \sigma^y
](\partial_{x_1}-\partial_{x'_1})-[\sigma^\mu,
\sigma^x](\partial_{y_1}-\partial_{y'_1}))_{\eta'\eta}
G_{\eta\nu\eta'\nu'}(1,2;1',3)\right\}_{1'=1^+},\end{eqnarray} where
$G_{\eta\nu\eta'\nu'}(1,2;1',2')=\langle\mathcal
{T}[\psi_{\eta}(1)\psi_{\nu}(2)\psi_{\eta'}^{\dag}(1')\psi_{\nu'}^{\dag}(2')]
\rangle$. Here we only focus our attention on the identities related
to spin transport, so, take the $\sigma^{\mu}=\sigma^{x}$,
$\sigma^{y}$ and $\sigma^{z}$ respectively and substitute
Eq.(\ref{correlation}) into (\ref{identity}), we get
\begin{eqnarray}\label{identity-x}\sigma^{x}_{\eta'\eta}\left\{\frac{\partial}{\partial
t_1}G_{\eta\nu\eta'\nu'}(1,2;1^+,3)+\nabla\cdot\frac{1}{2im}\left[\left
(\nabla_1-\nabla_{1'}\right)G_{\eta\nu\eta'\nu'}(1,2;1',3)\right]_{1'=1^+}\right\}
+\alpha\partial_{y_1}G_{\eta\nu\eta\nu'}(1,2;1^+,3)\nonumber\\
=i\alpha \sigma^z_{\eta'\eta}\left[(\partial_{x_1}-\partial_{x'_1})
G_{\eta\nu\eta'\nu'}(1,2;1',3)\right]_{1'=1^+}-
i\left[\delta(1-2)\sigma^{x}G(2,3)-\delta(1-3)G(2,3)\sigma^{x}\right]_{\nu\nu'}\end{eqnarray}
\begin{eqnarray}\label{identity-y}\sigma^{y}_{\eta'\eta}\left\{\frac{\partial}{\partial
t_1}G_{\eta\nu\eta'\nu'}(1,2;1^+,3)+\nabla\cdot\frac{1}{2im}\left[\left
(\nabla_1-\nabla_{1'}\right)G_{\eta\nu\eta'\nu'}(1,2;1',3)\right]_{1'=1^+}\right\}
-\alpha\partial_{x_1}G_{\eta\nu\eta\nu'}(1,2;1^+,3)\nonumber\\
=i\alpha \sigma^z_{\eta'\eta}\left[(\partial_{y_1}-\partial_{y'_1})
G_{\eta\nu\eta'\nu'}(1,2;1',3)\right]_{1'=1^+}-
i\left[\delta(1-2)\sigma^{y}G(2,3)-\delta(1-3)G(2,3)\sigma^{y}\right]_{\nu\nu'}\end{eqnarray}
\begin{eqnarray}\label{identity-z}&&\sigma^{z}_{\eta'\eta}\left\{\frac{\partial}{\partial
t_1}G_{\eta\nu\eta'\nu'}(1,2;1^+,3)+\nabla\cdot\frac{1}{2im}\left[\left
(\nabla_1-\nabla_{1'}\right)G_{\eta\nu\eta'\nu'}(1,2;1',3)\right]_{1'=1^+}\right\}
\nonumber\\
&&=-i\alpha\left\{[
\sigma^x_{\eta'\eta}(\partial_{x_1}-\partial_{x'_1})+\sigma^y_{\eta'\eta}(\partial_{y_1}-\partial_{y'_1})]
G_{\eta\nu\eta'\nu'}(1,2;1',3)\right\}_{1'=1^+}\nonumber\\&&~~
-i\left[\delta(1-2)\sigma^{z}G(2,3)-\delta(1-3)G(2,3)\sigma^{z}\right]_{\nu\nu'}\end{eqnarray}

If we define
$L_{\eta\nu\eta'\nu'}(1,2;1',2')=G_{\eta\nu\eta'\nu'}(1,2;1',2')-G_{\eta\eta'}(1,1')G_{\nu\nu'}(2,2')$,
and substitute this definition into Eq.(\ref{identity-x})
(\ref{identity-y}) and (\ref{identity-z}), we shall obtain similar
identities connecting $L$ and $G$, in which all
$G_{\eta\nu\eta'\nu'}(1,2;1',2')$'s are replaced by
$L_{\eta\nu\eta'\nu'}(1,2;1',2')$'s, while the contribution of the
second term $G_{\eta\eta'}(1,1')G_{\nu\nu'}(2,2')$ of $L$ will be
vanishing owing to the general continuity equation. The $L$ is a
general response function, which can also be given by
$$L_{\eta\nu\eta'\nu'}(1,2;1',2')=-\left[\frac{\delta
G_{\nu\nu'}(2,2';U)}{\delta U_{\eta'\eta}(1',1)}\right]_{U=0},$$
where the $U$ dependence of $G(U)$ is given by
$$iG_{\eta\eta'}(1,1';U)=\frac{\langle\mathcal
{T}[\psi_{\eta}(1)\psi_{\eta'}^{\dag}(1')e^{-i\int
U_{\lambda\lambda'}(\bar{2},\bar{2}')\psi^{\dag}_{\lambda}(\bar{2})\psi_{\lambda'}(\bar{2}')}]
\rangle}{\langle\mathcal {T}[e^{-i\int
U_{\lambda\lambda'}(\bar{2},\bar{2}')\psi^{\dag}_{\lambda}(\bar{2})\psi_{\lambda'}(\bar{2}')}]
\rangle}.$$

For the system with other kind of spin-orbit couplings, e.g.,
Luttinger model, Dresselhaus spin-orbit coupling or 2D cubic Rashba
model,\cite{Luttinger,Dresselhaus, Dyakonov, Rashba-Sov} the
Hamiltonian maybe include square or cubic terms in momentum.
According to the definition, the expressions of the spin-current and
torque will become more complex, so, the concrete form of these
identities will be also become more complex, but the principle is
the same.

\section{Summary}
Some exact identities connecting the correlation functions and the
one-particle Green's function in spin-orbit coupling systems were
derived. These identities are very similar to the usual Ward
identity in charge or particle transport theory, although the former
are originated from a more general continuity-like equation while
the latter is due to the conservation law or gauge invariance. These
identities can provide a criterion of whether an approximate
calculation of spin current and spin torque in the presence of
spin-orbit coupling is self-consistent. Therefore, in practical
calculations we must adopt some satisfactory approximations, which
can preserve these identities.

\begin{acknowledgments}
We are very thankful to Zhongshui Ma and Dingping Li for their
helpful suggestions. This work was supported by the Research Fund
for the Doctoral Program of Higher Education and the China
Postdoctoral Science foundation.
\end{acknowledgments}

\end{document}